# Student Certificate Sharing System Using Blockchain and NFTs

Prakhyat Khati[1][0009-0000-1383-888X], Ajay Kumar Shrestha[2][0000-0002-1081-7036] and Julita Vassileva[1][0000-0001-5050-3106]

[1] University of Saskatchewan, Saskatoon, SK S7N 5C9, Canada
`prakhyat.khati@usask.ca, jiv@cs.usask.ca`
[2] Vancouver Island University, Nanaimo, BC V9R 5S5, Canada
`ajay.shrestha@viu.ca`

**Abstract.** In this paper, we propose a certificate sharing system based on blockchain that gives students authority and control over their academic certificates. Our strategy involves developing blockchain-based NFT certifications that can be shared with institutions or employers using blockchain addresses. Students may access the data created by each individual institute in a single platform, filter the view of the relevant courses according to their requirements, and mint their certificate metadata as NFTs. This method provides accountability of access, comprehensive records that are permanently maintained in IPFS, and verifiable provenance for creating, distributing, and accessing certificates. It also makes it possible to share certificates more safely and efficiently. By incorporating trust factors through data provenance, our system provides a countermeasure against issues such as fake and duplicate certificates. It addresses the challenge of the traditional certificate verification processes, which are lengthy manual process. With this system, students can manage and validate their academic credentials from multiple institutions in one location while ensuring authenticity and confidentiality using digital signatures and hashing for data protection against unauthorized access. Overall, our suggested system ensures data safety, accountability, and confidentiality while offering a novel approach to certificate distribution.

## 1 Introduction

Over the past years, the use of blockchain technology has gained immense popularity across various domains. One of the areas where the application of blockchain technology has the potential to revolutionize is educational data sharing. Educational data such as academic certificates, and student transcript data are all important personal data which are unique as it identifies each student. The uniqueness of the data makes it fit to be represented as Non-fungible token (NFT). NFT is considered a piece of data stored on a blockchain. It certifies the uniqueness of an asset. NFTs are seen to be used to prove the authenticity, legitimacy of digital assets and real assets that have a digital footprint [1]. Numerous studies have explored the use of blockchain-based technologies for managing certificates. However, the utilization of Non-fungible Tokens (NFTs) in the domain of educational certificate management is still at an incipient stage.

In recent years, there has been a significant increase in the number of students studying abroad or applying for jobs in foreign countries. These students often need to provide proof of their academic credentials through certificates. However, the traditional paper-based certificate system requires multiple steps for certification, translation, and authentication, which can be time-consuming and costly. Additionally, students with multiple certificates must repeat the process for each certificate they hold, causing further delays and expenses. Also, the certificate once shared cannot be retrieved back, this could lead to misuse and data trust issues of such important credentials. To address certificate sharing, some institutions have implemented web2-based centralized systems for sharing certificates. However, they still face challenges such as a lack of global standardization and are often limited to a group of universities or institutions. Another major drawback of these approaches is that they lack data provenance, as the history and origin of the certificates cannot be traced in such implementations. This can lead to issues with trust and legitimacy when it comes to verifying the authenticity of certificates and the institute that issued the certificate. As a result, there is a growing need for a more secure and reliable global system for managing ownership and sharing academic certificates which has led to the exploration of blockchain-based solutions, such as the use of NFTs.

This paper proposes an NFT-based certificate sharing framework that provides an immutable and secure platform for academic institutions and students to manage ownership and share certificates. The framework leverages blockchain technology and smart contracts to ensure transparency, traceability, and authenticity of certificates. On top of that NFTs are used for assigning ownership, which makes it easier to track and verify the authenticity. By having metadata linked to them, NFTs can offer additional information about the certificate, including details such as who issued it, when it was issued, and a description of the achievement. This valuable information can be utilized to provide more context around the certificate, and to establish varying levels of access rights to the credentials, enabling individuals to grant or revoke access as necessary.

The rest of the paper is organized as follows. Section 2 describes the overview of blockchain, smart contracts and NFTs. A brief analysis of the existing architecture with their limitations is given in Section 3. After that Section 4 presents the solution architecture, our implemented model and describes our implementation details with the proposed approach of evaluation of the implemented system. Finally, the Section 5 concludes and summarizes the paper with future directions.

## 2 Background

The section provides background information about blockchain, smart contracts, non-fungible tokens (NFTs) and some of the terminology that is going to be used in the proposed framework.

Blockchain, also known as distributed ledger technology (DLT), is a data structure that allows for the creation of both private and public digital transactions by maintaining a shared ledger of transactions among network nodes [2]. The Ethereum community built an automation layer on top of a public permissionless blockchain using smart contracts managed by the decentralized network.

Smart contracts can be thought of as "if/then" conditional statements kept on the distributed ledger; they are pieces of executable code that are executed when triggered by an authorized or agreed-upon occurrence. The applications created on top of smart contracts are supported by the state-transition mechanism [3]. All participants share the states that contain the instructions and parameters, ensuring the accuracy of the directives.

Non-fungible tokens are distinct and non-interchangeable tokens stored on a blockchain. The standard for creating and maintaining non-fungible tokens using Ethereum smart contacts is governed by the ERC-721 standard. The standard defines how each NFT can represent someone's ownership of a specific digital and physical asset cryptographically. NFTs applications rely on blockchain technologies with smart contracts to ensure ownership, provenance, and exclusivity of the asset.

ERC-4361(SignIn with Ethereum) [4] is a standard proposed by the Ethereum community for enabling a decentralized authentication process with blockchain based wallets such as MetaMask. With this standard, users can sign in to decentralized applications (dApps) using their Ethereum wallets instead of creating separate login credentials for each dApp. The standard is based on the OAuth 2.0 framework and aims to provide a simple and secure way for dApps. Here we will be using MoralisSDK [5] to integrate with firebase, the backend, and the database. The centralized portion of our framework is handled using firebase and fire store.

IPFS (InterPlanetary File System) enables decentralized file sharing and storage [6]. It uses a peer-to-peer network of nodes to store and retrieve content, in contrast to conventional web protocols that depend on centralized servers. The unique content-addressing method used by IPFS allows files to be identified by their hash. This makes it possible to retrieve files even when the original source is unavailable, and it also makes file verification and versioning simple. It is possible to use hybrid encryption to prevent unauthorized access to stored content on IPFS. Here we have used pinata as a platform to store our NFTs metadata off-chain. These data are addressable using the hash, also known as Content-Identifiers (CIDs).

## 3 Related Works

Blockchain technology has the potential to revolutionize the way we issue and verify certificates and credentials. In the early stages of blockchain, various institutes presented several platforms and research solutions for generating certificates and badges. These platforms allowed universities to issue and backup certificates for their students or supported self-generated certificates and badges within their own private blockchain ecosystem. NFTs were not utilized by any of these applications. The author in [7] pre-sented a detailed comparison of what are the benefits and challenges faced by the existing blockchain based education certificate sharing systems and proposed a new framework called NFTCert [7].

In late 2021 and early 2022, we saw the development of using NFTs for managing and defining ownership of digital arts. This spiked the exploration of the potential benefits of using NFTs for adding uniqueness to certificates, and badges to define ownership. Some proposed certificate systems, such as NFTCert [7] and "Ethernal Digital Certificates"[8] either incorporated NFTs to their existing blockchain frameworks or proposed new ones to solve problems in the existing ones.

NFTCert is a platform that allows the institute to create their student NFT-based certificates and transfer the ownership to the student. It also provides hash value for verifying the authenticity of NFT-based certificates. The control over data is only accessible to the owner who mints the NFT, here in this case it's the University. The platform also incorporated blockchain Oracle to add an online payment gateway for the necessary payment to retrieve the certificate.

While these NFT-based systems have played a role in reducing fraudulent activities by taking advantage of the offered technologies, they still have several limitations. For example, in NFTCert, the institute only has the right to mint the NFT. From the student point of view, this approach is time consuming as it involves more interaction with the institute, and the paper does not clearly state whether the ownership or the right to access the NFT is transferred to the student address; this make the student completely dependent upon the institute Also, here student cannot create purpose centric view of certificates, for example, if a student is applying for a software developer job, not all subjects or credits that he studied might be relevant to that particular job, in such as case, it would have been convenient if the student could be able to give access only to certain subjects that are relevant. However, this should not compromise the authenticity of the credential issued by the university. Furthermore, this platform focuses on NFT certificate management rather than sharing the NFT certificate and is centered towards private network implementation. This may limit the scope of the system and may not meet the needs of all stakeholders.

In contrast to these approaches, in this paper, we propose a novel NFT-based certificate-sharing system which not only overcomes the weaknesses of these systems but also introduces new features for managing and sharing NFTs.

## 4      Solution Framework and Discussions

The proposed solution framework for certificate issuance and ownership offers a decentralized approach that grants students maximum ownership of their certificates. Under this framework, university administrators can register their institutions and create university and student profiles containing metadata pertaining to the issuer address and signature for verification purposes. This metadata is then shown to the student's dashboard, which they can access by logging into the system using decentralized authentication via a MetaMask wallet.

Decentralized authentication involves using the Ethereum account to authenticate with an off-chain service by signing a standardized message format parameterized by scope, session detail, and security mechanisms. This can play a crucial role in the context of NFT certificate and credit sharing systems that provide ownership to students and implement access control. By leveraging the Ethereum blockchain, decentralized authentication can enable students to create self-sovereign identities using decentralized identity (DID) protocols like Decentralized Identifiers (DIDs). This can help to ensure that students have full control over their personal information and sole access to their certificates and credit information which is only granted to authorized parties.

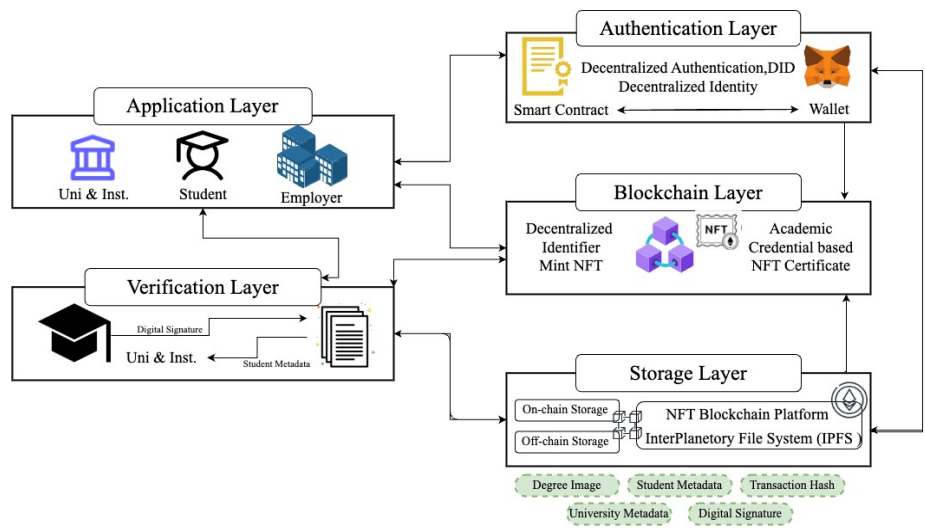

**Fig. 1.** A general system overview

With reference to this proposed general system overview in Fig. 1., the actual implementation is portrayed with the solution framework in Fig. 2. Here, the University generates a hash of the metadata and signs the hash using its private key to create a digital signature. The digital signature is then added to the metadata along with the original metadata. The university creates the NFT metadata, mints and then transfers the NFT to the student wallet giving the student full ownership of the NFT.

Students have the flexibility to decide how many viewNFTs they want to mint from their original certificate. ViewNFT is a novel term introduced in this framework to reference the NFT that are created by students. As per the smart contract rules, these viewNFTs will have flag that enables student to filter only courses and does not allow to alter any other information of the credentials. So now a student can just put relevant courses as a view transcript with degree completion certificate and mint these data as viewNFT certificate. The minted viewNFT will have student as the owner and the student can then share and allows access to these certificates and credential wherever necessary. Most importantly, even after sharing the certificate and transcript, the ownership remains with the student due to the rules set in the smart contract. This allows the student to take back ownership of their certificate even after sharing it with others. The NFT can be shared using the receiver's wallet address, and the student can fetch the NFT back to their own dashboard by changing the access rights.

Upon receiving the viewNFT generated by the student, the receiver will receive a notice mentioning that it is a viewNFT, and the student chose to hide certain subjects in this viewNFT. The receiver can use the university public key to decrypt the digital signature attached to the viewNFT, the hash obtained afterwards represents the original NFT from where the viewNFT is made. The function verifies the ownership and authenticity by comparing the hashes of the original data attached to the viewNFT and other supporting elements in the metadata.

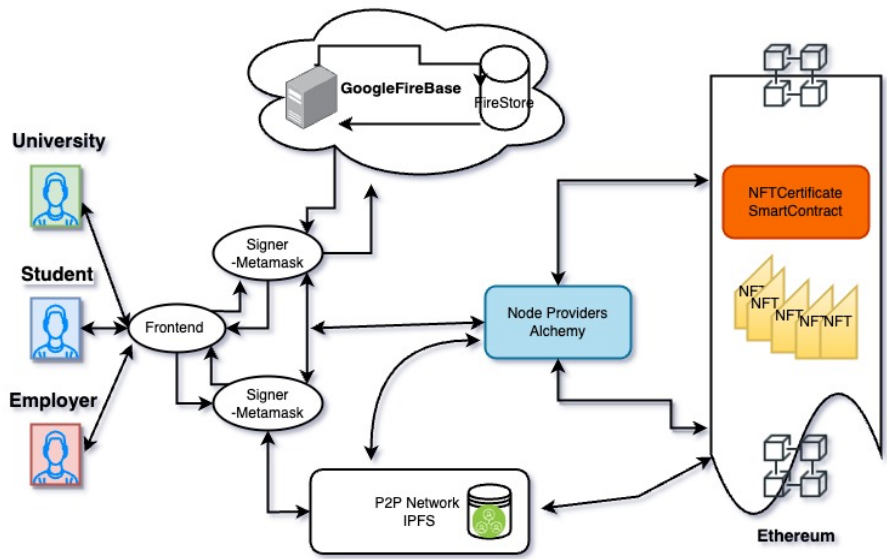

**Fig. 2.** System architecture

In summary, the solution framework grants students full ownership of their certificates, simplifies the transaction process, provides greater control and customization options, and offers an added layer of security to ensure that the student's data always remains under their control. The framework utilizes decentralized authentication, allowing students to access their profile information without the need for a centralized authority. University administrators play a crucial role in registering their institutions and creating student profiles, while employers can view and verify the certificates issued to potential employees.

### 4.1 Schema

According to the ERC721 standard, we define the metadata JSON schema to represent the certificate. It includes institute details, degree details, student details, course issued to the students and the respective wallet address details as shown in Fig. 3.

```
{
"Title": "Certificate Metadata",
"Description": {                    "OriginalNFT": {
  "Creator": String,                  "TokenId": uint256,
  "Owner": String,                    "TokenURI": String,
  "StudentID": String,                "Subjects": [{"mark": String,
  "StudentWalletAddr": String,        "subject": String},
  "StudentName": String,              {"mark": String, "subject":
"StudentDescription": String,       String}]},

                                    "UniversityName": String,
"ViewNFT": {                        "UniversityID": String,
"TokenId": uint256,                 "UniversityAddress":
"TokenURI": String,                 String,
"ViewSubjects": [                   "UniversityMetadata":
        {"mark": String,            String,
"subject": String},                 "Degree": String,
  {"mark": String, "subject":       "CertificateHash": String
String}
      ]}}
```

**Fig. 3.** ViewNFT certificate metadata

Here, the course in the schema is dynamically set by the university/institute that is creating the profile for the student. The viewSubjects are the subjects that students filtered from the original subject to create a viewNFT. The smart contract is deployed just once for each node on the Ethereum blockchain which stores the access control and tokenID to identify the user who minted their certificate issued to them by their University/Institute. Here the receivers can view the course subject that the student has set, and the smart contract keeps track of the changes made by the student. If the receiver wants to verify the authenticity of the academic credential, the receiver can view the original metadata validated by the university and decrypt the signature using the university public key to get access to the hash. After receiving the hash, the receiver can verify the authenticity of the data by comparing the hash value. Since the NFT allows verification of its history regarding ownership, it is easier to verify the owner's address.

We have used OpenZeppline smart contracts to create our NFTCertificate sharing smart contact. The smart contract developed with Solidity contains the following functions.

```
Contract NFTCertificate is ERC721URIStorage {
                                    Struct TokenTransferScheduler
Struct ListedToken                  {
{                                   uint256 transferBackTime
uint256 tokenId,                    address transferBackTo
address payable owner,              }
address payable viewer,             Event TokenListSuccess
uint256 transferBackTime,           (
bool currentlyListed,               uint256 indexed tokenId,
bool currentViewer                  address owner,
}                                   address viewer,
                                    uint256 transferBackTime,
```

```
                                    bool currentlyListed,
                                    bool currentViewer )

function createToken (string memory tokenURI, address Studen-
tAddress) public payable returns (uint)
function createListedToken (uint256 tokenId) private
function getAllNFTs () public view returns (ListedToken []
memory)
function getMyNFTs () public view returns (ListedToken []
memory)
function getTrasnferedNFTs () public view returns (ListedToken
[] memory)
function executeTransfer (uint256 tokenId, address receiver,
uint256 transferBackTime) onlyOwner
function transferOwnership (uint256 tokenId) onlyOwner
```

**Fig. 4.** Smart contract of NFT based certificate sharing system.

Here, the createToken () method takes the tokenURI as parameter, the tokenURI is hash value generated after uploading the metadata to IPFS. The data from the Firestore is given in JSON format with all the necessary metadata. The metadata is first uploaded to IPFS using Pinata API. This function mints an NFT and assigns ownership to the student wallet address. The CreateListedToken () function is used solely to keep track of the number of NFTs issued via the smart contract. To display data on the frontend of the application, we need to create additional methods, such as getAllNFTs (), which lists all NFTs issued using the smart contract, and viewNFT (), which displays a particular NFT. Similarly, the getMyNFTs () method returns a list of all the viewNFTs owned by the current user, while the getTransferredNFTs () method retrieves a list of viewNFTs that have been transferred to universities and employers. The executeTransfer () method in the smart contract is used to transfer view rights to an Ethereum address. The owner can set a default time period after which the transferOwnership function is executable to reclaim the viewing rights of the NFTs. The owner can revoke view rights by calling the transferOwnership () method from the application frontend.

### 4.2 Performance Evaluation

This section covers the evaluation of the performance of the smart contract. The evaluation focuses on two aspects: transaction latency and resource consumption. The network setup used by default on Kaleido consists of two nodes, each running on 0.5vCPU with 1024MB of memory. Fig. 5. shows the CPU resource consumption of both nodes and the system. The transaction latency recorded was 4.16 seconds. Here, y-axis of the graph represents the percentage of CPU usage, while the x-axis represents the duration.

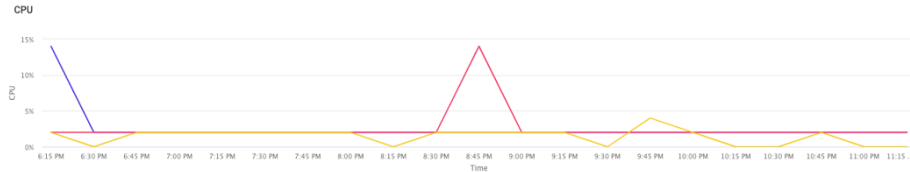

**Fig. 5.** CPU utilization of nodes

During the experiment, the nodes consumed only 2% of the CPU to maintain operations. However, there was a temporary increase in CPU usage for a few minutes, up to 14%, when we triggered the minting of an NFTCertificate using the deployed smart contract.

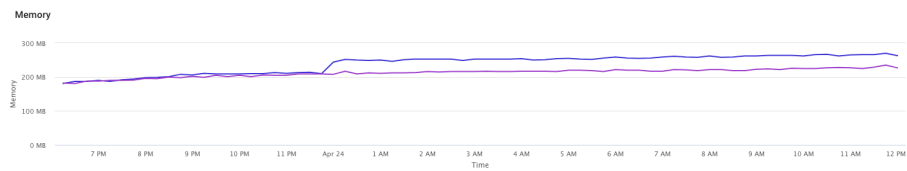

**Fig. 6.** Memory used of nodes.

In Fig.6., the value on the y-axis denotes memory usage in MB and the x-axis denote the time in minutes within the day. The data indicates that each node utilized less than 253 MB of memory. At 12.15 AM on the test day, Node 2 on the graph used the smallest amount of memory, which was recorded as 227 MB. On the other hand, Node 1 used the largest amount of memory, approximately 252 MB, at around the same time. The memory consumption pattern remained steady throughout the day with no significant spikes observed. Finally, Fig. 7. illustrates the number of transactions and blocks per hour. On average, there were 0.02 transactions (Tx) and 11 blocks per hour. It took an average of 5 seconds for a new block to be added to the blockchain.

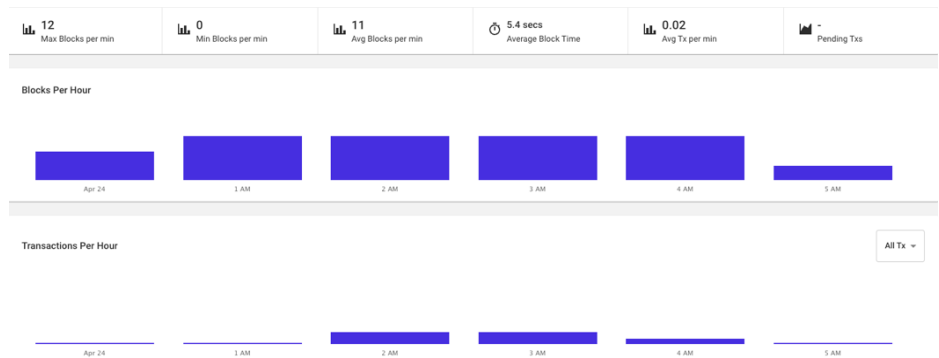

**Fig. 7.** Disk utilization of nodes

## 5   Conclusion

In summary, our proposed solution framework provides a decentralized approach to certificate issuance and ownership, giving students full control over their credentials. The use of Ethereum currency for transactions simplifies the process, while the ability to mint viewNFT from certificate credentials allows for greater customization and control over their academic certificate. Our future work involves improving the current model by studying users' attitudes to academic certificate sharing with NFT and blockchain where students have ownership and access control even after sharing their degree. We are also planning to evaluate the usability, trust perceived security, usefulness of the proposed framework using the validated constructs of the Technology Acceptance Model (TAM). We will also be investigating the possibility of introducing incentive for universities through this framework.